\documentstyle[12pt]{article}
\begin{document}
\newcommand\beq{\begin{equation}}
\newcommand\eeq{\end{equation}}
\newcommand\bea{\begin{eqnarray}}
\newcommand\eea{\end{eqnarray}}
\begin{center}
{\Large A DMRG Study of Low-Energy Excitations and Low-Temperature 
Properties of Alternating Spin Systems}
\end{center}

\centerline{ \bf Swapan K. Pati$^{1}$, S. Ramasesha$^{1,3}$ 
and Diptiman Sen$^{2,3}$} 

\centerline{\it $^1$ Solid State and Structural Chemistry Unit} 
\centerline{\it $^2$ Centre for Theoretical Studies}  
\centerline{\it Indian Institute of Science, Bangalore 560012, India} 
\centerline{\it $^3$ Jawaharlal Nehru Centre for Advanced Scientific Research} 
\centerline{\it Jakkur Campus, Bangalore 560064, India}

\begin{abstract}

We use the density matrix renormalization group (DMRG) method to study the 
ground and low-lying excited states of three kinds of uniform and dimerized 
alternating spin chains. The DMRG procedure is also employed to obtain 
low-temperature thermodynamic properties of these systems. We consider a $2N$ 
site system with spins $s_1$ and $s_2$ alternating from site to site and 
interacting via a Heisenberg antiferromagnetic exchange. The three systems 
studied correspond to $(s_1 ,s_2 )$ being equal to $(1,\frac{1}{2})$, 
$(\frac{3}{2},\frac{1}{2})$ and $(\frac{3}{2},1)$; all of them have very 
similar properties. The ground state is found to be ferrimagnetic with total 
spin $s_G =N(s_1 - s_2)$. We find that there is a gapless excitation to a 
state with spin $s_G -1$, and a gapped excitation to a state with spin 
$s_G +1$. Surprisingly, the correlation length in the ground state is found to 
be very small for this gapless system. The DMRG analysis shows that the chain 
is susceptible to a conditional spin-Peierls instability. Furthermore, our 
studies of the magnetization, magnetic susceptibility $\chi$ and specific heat 
show strong magnetic-field dependences. The product $\chi T$ shows a minimum 
as a function of temperature $T$ at low magnetic fields; the minimum vanishes 
at high magnetic fields. This low-field behavior is in agreement with earlier 
experimental observations. The specific heat shows a maximum as a function of 
temperature, and the height of the maximum increases sharply at high magnetic 
fields. Although all the three systems show qualitatively similar behavior, 
there are some notable quantitative differences between the systems in which
the site spin difference, $|s_1 - s_2|$, is large and small respectively.

\end{abstract}

\newpage

\section{Introduction}

There have been a number of experimental efforts over the last several
years to synthesize molecular systems showing spontaneous magnetization
\cite{flem,alle,mill}. These studies have led to the recent 
realization of systems such as
$NPNN$ ($para$-nitrophenyl nitronyl nitroxide) and $C_{60}-TDAE$ (Tetrakis 
Dimethyl Amino Ethylene) which show ferromagnetic to paramagnetic
transition at low-temperatures. There also exist theoretical models
\cite{ovch,sinha} that predict ferrimagnetism in organic polymer systems, but 
a truly extended organic ferrimagnet has not been synthesized although an
oligomer with a ground state spin of $S=9$ is now known. On the
inorganic front, pursuit of molecular magnetism has been vigorous,
and there has been success in the synthesis of molecular systems
showing spontaneous magnetization at low-temperatures\cite{stein76,kahn1}. 
These are quasi-one-dimensional bimetallic molecular magnets in which each 
unit cell contains two spins with different spin values \cite{kahn1}.
These systems contain two transition metal ions per unit cell, with
the general formula $ACu(pbaOH)(H_2O)_3.2H_2O$, where $pbaOH$ is
2-hydroxo-1,3-propylenebis(oxamato) and $A$=$Mn$, $Fe$, $Co$, $Ni$, and they
belong to the alternating or mixed spin chain family \cite{kahn2}. These
alternating spin compounds have been seen to exhibit ferrimagnetic behavior.
It has been possible to vary the spin at each site from low values where
quantum effects dominate to large values which are almost classical.

There have been a number of theoretical investigations for 
quantum ferrimagnetic systems in recent years \cite{vega,alcaraz}. However,
these models consider complicated multispin interactions, while 
very little is known about the simple Heisenberg model with purely quadratic
interactions. There have been some analytic studies of alternating
spin-1/spin-$\frac{1}{2}$ system \cite{mik} and a detailed spin-wave
analysis followed by a DMRG study \cite{pati} corresponding to the 
$Cu-Ni$ bimetallic chain with the simple Heisenberg model. These studies 
confirm that these ferrimagnetic 
systems can be accurately described by a pure Heisenberg spin model.

Theoretical studies of alternating spin systems have so far
been concerned only about the spin-1/spin-$\frac{1}{2}$ bimetallic
chains; further, the thermodynamic properties have not been
explored in detail. The thermodynamic behavior of these alternating
spin compounds is very interesting \cite{kahn2,kahn3}. In very low magnetic 
fields, these systems show one-dimensional ferrimagnetic behavior. The 
$\chi T$ vs. $T$ (where $\chi$ is the magnetic susceptibility and $T$ the
temperature) plots show a rounded minimum. As the temperature is increased,
$\chi T$ decreases sharply, goes through a minimum before increasing 
gradually. The temperature at which this minimum occurs differs from
system to system and depends on the site spins of the chain.
The variation of the field induced magnetization with temperature is also
interesting as the ground state is a magnetic state. At low magnetic fields,
the magnetization decreases with increasing temperature at low temperatures.
However, at moderate magnetic
fields, with increase in temperature, the magnetization slowly increases,
shows a broad peak and then decreases. Such a behavior has been studied 
theoretically by us for the spin-1/spin-$\frac{1}{2}$ system \cite{pati}.

These interesting observations have motivated us to study of ferrimagentic 
systems with arbitrary spins $s_1$ and $s_2$ alternating from site to site. 
It would be quite interesting to 
know the thermodynamic properties of these systems with varying $s_1$ 
and $s_2$. In the previous paper, we predicted some interesting features 
of the thermodynamic properties of the spin-1/spin-$\frac{1}{2}$
system at high magnetic fields. We focus on the high field behavior of the
general ferrimagnetic chains to find whether the observed properties are 
generic to these systems or are dependent on specific $s_{1}$ and $s_{2}$ 
values.

In this paper, we study the low-lying excited states and 
low-temperature properties of the spin-$\frac{3}{2}$/spin-$1$ and
spin-$\frac{3}{2}$/spin-$\frac{1}{2}$ chains and rings and compare them with
with those of the corresponding spin-$1$/spin-$\frac{1}{2}$ systems. 
We have employed the density matrix renormalization group (DMRG)
method which has proved to be the best numerical tool for 
low-dimensional spin systems in recent years \cite{white1}. The ground state 
energy per site, the spin excitation gap and the two-spin correlation 
functions obtained from this method have been found to be accurate to several
decimal places \cite{white2,hall,chitra} when compared with Bethe-ansatz
results (where possible) and exact diagonalization results of small
systems. In the DMRG method, spin parity symmetry can be used
to characterize the spin states along with the $S^z_{tot}$ as the good
quantum numbers. The DMRG calculations have been carried out on
chains and rings with alternate spin-$s_1$/spin-$s_2$ sites, with $s_1$ fixed
at $\frac{3}{2}$ and $s_2$ being $1$ or $\frac{1}{2}$. Studies of the 
ground state and low-lying excited states are reported in detail and
compared with those of $s_1=1$ and $s_2=\frac{1}{2}$ system. Furthermore, by
resorting to a full diagonalization of the DMRG Hamiltonian matrix in
different $S_z$ sectors, we have also obtained the low-temperature
thermodynamic properties of these systems. The theromodynamic properties
we discuss will include low and high field magnetization, magnetic
susceptibilty and specific heat, all at low temperatures. 

The paper is organized as follows. In section 2, we present properties of the 
ground and low-lying excited states and compare them with the results of spin 
wave theory. In section 3, we discuss the low-temperature thermodynamic 
properties of the systems. We summarize our results in the last section.
 
\section{Ground State and Excitation Spectrum}

We start our discussion with the Hamiltonian for a chain with spins $s_1$ and
$s_2$ on alternating sites (with $s_1 > s_2$, without loss of generality).
\beq
H ~=~ J ~\sum_{n} ~[~ (1+\delta) ~{\vec S}_{1,n} \cdot {\vec S}_{2,n} ~+~
(1-\delta) ~{\vec S}_{2,n} \cdot {\vec S}_{1,n+1} ~]~,
\label{ham}
\eeq
where the total number of sites is $2N$ and the sum is over the
total number of unit cells $N$. ${\vec S}_{i,n}$ corresponds to the spin
operator for the site spin $s_i$ in the $n$-th unit cell. The 
exchange integral $J$ is taken to be positive for all our calculations;
$\delta$ is the dimerization parameter and it lies in the range $[0,1]$.

\subsection{Summary of results from spin-wave theory}

Before describing our numerical results, we briefly summarize the results
of a spin wave analysis for the purposes of comparison \cite{pati}. We will 
first state the results with $\delta =0$. According to spin wave theory, the 
ground state has total spin $s_G =N(s_1 -s_2 )$. Let us define a function
\beq
\omega(k) ~=~ J ~\sqrt{(s_1 -s_2)^2 + 4 s_1 s_2 \sin^2 (k/2)} ~,
\eeq
where $k$ denotes the wave number. Then the ground state energy per site
is given by
\beq
\epsilon_0 ~=~ \frac{E_0}{2N} ~=~ - ~Js_1 s_2 ~+~ \frac{1}{2} ~\int_0^{\pi} ~
\frac{dk}{\pi} ~[~ -~J (s_1 +s_2 ) ~+~ \omega(k) ~]~.
\eeq
The lowest branch of excitations is to states with spin $s=s_G -1$, with 
the dispersion
\beq
\omega_{1}(k) ~=~ J ~(-s_1 + s_2) ~+~ \omega(k) ~;
\eeq
the gap vanishes at $k=0$. There is a gapped branch of excitations to
states with spin $s=s_G +1$, with the dispersion
\beq
\omega_{2}(k) ~=~ J ~(s_1 - s_2) ~+~ \omega(k) ~;
\eeq
the minimum gap occurs at $k=0$ and is given by $\Delta = 2J (s_1 - s_2 )$.
In the ground state with $S_z = s_G$, the sublattice magnetizations are given 
by the expectation values
\bea
<S^z_{1,n}> &=& (~s_1 +\frac{1}{2}~) ~-~ \frac{1}{2} ~\int_0^{\pi} ~
\frac{dk}{\pi} ~\frac{J(s_1+s_2)}\omega(k) ~, \nonumber \\
<S^z_{2,n}> &=& s_1 ~-~ s_2 ~-~ <S^z_{1,n}> ~.
\eea
The various two-spin correlation functions decay exponentially with distance;
the inverse correlation length is given by $\xi^{-1} = \ln (s_1 /s_2 )$. The 
results with dimerization ($\delta > 0$) are very similar. In fact, within 
spin wave theory, the minimum gap $\Delta$ to states with spin $s=s_G +1$ is 
independent of $\delta$.

\subsection{Results from DMRG studies}

We have studied the system defined by equation (\ref{ham}) both with and 
without dimerization, $\delta \ne 0$ and $\delta =0$ respectively.
We study alternating spin-$\frac{3}{2}$/spin-$1$ (hereafter
designated as ($\frac{3}{2},1$)) and spin-$\frac{3}{2}$/spin-$\frac{1}{2}$ 
(to be called ($\frac{3}{2},\frac{1}{2}$)) chains with open boundary 
condition for the Hamiltonian (\ref{ham}) by employing the DMRG method. We 
compute the ground state properties for both the systems by studying chains of 
upto $80$ to $100$ sites. The number of dominant density matrix eigenstates,
$m$, that we have retained at each DMRG iteration also varies between 
$80$ to $100$ for both the systems. With the increase of the Fock space
dimensionality of the site spins, we increase $m$ to obtain more accurate
results. The DMRG procedure follows the usual steps
for chains discussed in earlier papers \cite{white1,hall,chitra}, except that 
the alternating chains studied here are not symmetric between the left and 
right halves; hence the density matrices for these two halves have to be 
constructed at every iteration of the calculations. We have verified the 
convergence of our results by varying the values of $m$ and the system size. 
The ground states of both the system lie in the $S_z =N(s_1 -s_2)$ sector,
as verified from extensive checks carried out by
obtaining the low-energy eigenstates in different $S_z$ sectors of a 
$20$-site chain. A state corresponding to the lowest energy in 
$S_z =N(s_1 -s_2)$ is found in all subspaces with $\vert S_z \vert \le 
N(s_1 -s_2)$, and is absent in subspaces with $\vert S_z \vert > N(s_1 -s_2)$. 
This shows that the spin in the ground state is $s_G =N(s_1 -s_2)$.

In a previous paper, we studied the alternating spin chain made up of
spin-$1$ and spin-${\frac{1}{2}}$ (hereafter refered as ($1,\frac{1}{2}$)) 
at alternate sites \cite{pati}. We compare here the results for all the 
three systems, nemely, ($\frac{3}{2},1$), ($\frac{3}{2},1/2$) and 
($1,\frac{1}{2}$) spin systems.
In figure 1, we show the extrapolation of the energy per site as a function 
of inverse system size for all three systems. The ground state energy per
site $\epsilon_0$ extrapolates to $-1.93096J$ for the ($\frac{3}{2},1$) system,
to $-0.98362J$ for ($\frac{3}{2},\frac{1}{2}$), and to $-0.72704J$ for 
($1,\frac{1}{2}$).  The spin wave analysis gives the values $-1.914J$ 
for ($\frac{3}{2},1$), $-0.979J$ for ($\frac{3}{2},\frac{1}{2}$) and $-0.718J$ 
for ($1,\frac{1}{2}$) which are all higher than the DMRG values.  
It is interesting to note that, in the alternating spin cases, the energy 
per site lies in between the values for the pure spin-$s_1$ 
uniform chain and the pure spin-{$s_2$} uniform chain.

In figure 2, we show the expectation value of site spin operator $S^z_{i,n}$ 
( spin density ) at all the sites for the ($\frac{3}{2},1$), 
($\frac{3}{2},\frac{1}{2}$) and ($1,\frac{1}{2}$) chains. The spin densities 
are uniform on each of the sublattices in the chain for all the three systems. 
For the ($\frac{3}{2},1$) chain, the spin density at a spin-$\frac{3}{2}$ 
is $1.14427$ (the classical value is $\frac{3}{2}$), while, 
at a spin-$1$ site it is $-0.64427$ (classical value $1$). 
For the ($\frac{3}{2},\frac{1}{2}$)
case, the spin density at a spin-$\frac{3}{2}$ site is $1.35742$ and at a 
spin-$\frac{1}{2}$ site it is $-0.35742$. For the ($1,\frac{1}{2}$) case, the 
value at a spin-$1$ site is $0.79248$ and at a spin-$\frac{1}{2}$ site it is 
$-0.29248$. These can be compared with the spin wave values of $1.040$ and 
$-0.540$; $1.314$ and $-0.314$; and $0.695$ and $-0.195$ for the 
spin-$s_1$ and spin-$s_2$ sites of the ($\frac{3}{2},1$);
($\frac{3}{2},\frac{1}{2}$); and ($1,\frac{1}{2}$) systems 
respectively. We note that the spin wave analysis overestimates the quantum 
fluctuations in case of systems with small site spin values. We also notice 
that there is a greater quantum fluctuation when the difference
in site spin $|s_1 -s_2|$ is larger. This is also seen in spin-wave theory. 
The spin density distribution in an alternating ($s_1,s_2$) chain behaves 
more like that in a ferromagnetic chain rather than like an antiferromagnet, 
with the net spin of each unit cell perfectly aligned (but with small quantum
fluctuations on the individual sublattices). In a ferromagnetic ground 
state, the spin density at each site has the classical value appropriate
to the site spin, whereas for an antiferromagnet, this averages out to zero 
at each site as the ground state is nonmagnetic. From this viewpoint, the 
ferrimagnet is similar to a ferromagnet and is quite unlike an 
antiferromagnet. The spin wave analysis also yields the same physical picture.

Because of the alternation of 
spin-$s_1$ and spin-$s_2$ sites along the chain, one has to distinguish 
between three different types of pair correlations, namely, $< S^z_{1,0} 
S^z_{1,n}>$, $< S^z_{2,0} S^z_{2,n}>$ and $< S^z_{1,0} S^z_{2,n}>$.
We calculate all the three correlation functions with the mean values
subtracted out, since the mean values are nonzero in all these three
systems unlike in pure antiferromagnetic spin chains. In the DMRG procedure,
we have computed these correlation functions from the sites inserted at the 
last iteration, to minimize numerical errors. In figure 3, we plot the two-spin 
correlation functions in the ground state as a function of the distance
between the spins for an open chain of $100$ sites for all three cases. 
All three correlation functions decay rapidly with distance for each of
the three systems. From the figure it is clear that, except
for the $< S^z_{1,0} S^z_{2,n}>$ correlation, all
other correlations are almost zero even for the shortest possible distances. 
The $< S^z_{1,0} S^z_{2,n}>$ correlation
has an appreciable value [$-0.2$ for 
($\frac{3}{2},1$), $-0.07$ for ($\frac{3}{2},\frac{1}{2}$) and 
$-0.094$ for ($1,\frac{1}{2}$)] only for the nearest neighbors. 
This rapid decay of the correlation functions makes it difficult
to find the exact correlation length $\xi$ for a lattice model, 
although it is clear that $\xi$ is very small (less than one unit cell)
for the ($\frac{3}{2},\frac{1}{2}$) and ($1,\frac{1}{2}$) cases, and a 
little greater ($1<\xi <2$)
for the ($\frac{3}{2},1$) system. Spin wave theory gives $\xi =2.47$ for 
($\frac{3}{2},1$), $\xi =0.91$ for ($\frac{3}{2},\frac{1}{2}$), and 
$\xi =1.44$ for ($1,\frac{1}{2}$) cases. 

The lowest spin excitation of all the three chains is to a state with $s=s_G 
-1$. To study this state, we target the $2^{nd}$ state in the $S_z =s_G -1$ 
sector of the chain. To confirm that this state is a $s=s_G -1$
state, we have computed the $2$-nd state in $S_z =0$ sector and find that
it also has the same energy. However, the corresponding state is
absent in $S_z$ sectors with $\vert S_z \vert > s_G -1 $. Besides, from exact 
diagonalization of all the states of all the $s_1 -s_2$ alternating spin
chains with $8$ sites, we find that the energy orderings of the states is 
such that the lowest excitation is to a state with spin $s=s_G -1$. We have 
obtained the excitation gaps for all the three alternating spin chains in the
limit of infinite chain length by extrapolating from the plot of spin gap
vs. the inverse of the chain length (figure 4). We find that this 
excitation is gapless in the infinite chain limit for all three cases.
This property is quite unusual; the spin systems studied till date 
with antiferromagnetic exchage interactions have a small correlation length 
if and only if there is a finite gap to the lowest excited state of the 
system, from the ground state. Thus, this gapless excitation coexisting with 
very short correlation length in the ground state can indeed be taken as a 
signature of systems with spontaneous magnetization.

To characterize the lowest spin excitations completely,
we also have computed the energy of the $s=s_G +1$ state
by targetting the lowest state in the $S_z=s_G +1$ sector. In figure 5,
we plot the excitation gaps to the $s=s_G +1$ state from the ground state
for all three systems as a function of the inverse of the chain length. 
The gap saturates to a finite value of $(1.0221 \pm 0.0001)J$ for the 
($\frac{3}{2},1$) case, $(1.8558 \pm 
0.0001)J$ for ($\frac{3}{2},\frac{1}{2}$), and $(1.2795 \pm 0.0001)J$ 
for ($1,\frac{1}{2}$). It appears that the gap is also higher when the 
difference in site spins $|s_1 -s_2|$, is larger. The site spin densities 
expectation values computed in this state for all three cases are found 
to be uniform (i.e., independent of the site) on each of the sublattices.
This leads us to believe that this excitation cannot be characterized 
as the states of a magnon confined in a box, as has been observed for a 
spin-$1$ chain in the Haldane phase \cite{white2}. 

Earlier studies on pure spin-$1$ and pure spin-$\frac{1}{2}$ chains 
\cite{sumit} have revealed that with the 
alternation $\delta$ in the exchange parameter, the half-odd-integer 
spin chain will have an unconditional
spin-Peierls transition while for integer spin chains, the transition is
conditional. This conclusion has been drawn from the fact that, with the 
inclusion of $\delta$, the magnetic energy gain $\Delta E$ can be defined as
\beq
\Delta E(2N,\delta) ~=~ \frac{1}{2N} ~[~ E(2N,\delta) - E(2N,0) ~]~,
\eeq
where $E(2N,\delta)$ is the ground state energy of the $2N$-site system 
with an alternation $\delta$ in the exchange integral, and $E(2N,0)$ is the 
ground state energy of the uniform chain of $2N$ sites. For the pure spin 
chain, if we assume that $\Delta E$ varies as $\delta^{\nu}$ for small 
$\delta$, we find that $\nu < 2$ for the half-odd-integer spin chains
and $\nu=2$ for the integer spin chains \cite{sumit}.
Thus, for a half-odd-integer spin 
chain, the stabilization energy always overcomes the elastic energy,
whereas for the integer spin case, it depends on the lattice stiffness.

We have used DMRG calculations extend these studies to the alternating
spin systems. We obtain $\Delta E(2N,\delta)$, for small values of $\delta$ 
for all the three alternating spin chains. To determine the exact
functional form of the magnetic energy gain, we varied the chain length
from $50$ sites to $100$ sites and also $m$ (the number of states retained
in each DMRG iteration) from $80$ to $100$ to check the convergence of
$\Delta E$ with chain length.  We note that as we approach the 
classical spin limit, the convergence
is reached faster. The dependence of $\Delta E(2N,\delta)$ on
$1/2N$ is linear for all three cases for the $\delta$ values we have 
studied. Figure 6 gives a sample variation of $\Delta E(2N, \delta)$ with 
$1/2N$ for all the systems we have studied. This
allows us to extrapolate $\Delta E(2N,\delta)$ to the infinite chain
limit reliably for all the cases. In figure 7, we show the plot of 
$\Delta E(2N,\delta)$ vs. $\delta$ for finite $2N$ values and also the 
extrapolated infinite chain values for all three systems.
We see that there is a gain in magnetic energy upon dimerization even in the
infinite chain limit for all the systems. To obtain the exponent 
$\nu$, we plot $\ln \Delta E(2N,\delta)$ vs. $\ln \delta$ for the infinite
chain (figure 8). From this figure, we find that
in the alternating spin case, for the infinite chain 
$\Delta E \approx \delta^{2.00 \pm 0.01}$ for all three cases.
Thus the spin-Peierls transition appears to be close to being conditional 
in these systems. The magnetic energy gain per site for finite chains is 
larger than that of the infinite chain for any value of $\delta$ (figure 7). 
It is possible that the distortion in a {\it finite} chain is unconditional 
while that of the infinite chain is conditional for ferrimagnetic systems.

We have also studied the spin excitations in the dimerized alternating
($s_1,s_2$) chains, defined in equation (\ref{ham}).
We calculate the lowest spin excitation to the $s=s_G -1$ state
from the ground state. We find that the $s=s_G -1$ state is gapless from 
the ground state for all values of $\delta$. This result agrees with the spin 
wave analysis of the general ($s_1,s_2$) chain. The systems remain gapless 
even while dimerized unlike the pure antiferromagnetic dimerized spin chains. 
There is a smooth increase of the spin 
excitation gap from the ground state to the $s=s_G +1$ state with increasing 
$\delta$ for all three systems studied here. We have plotted this gap 
vs. $\delta$ in figure 9. The gap shows almost a linear behavior as a 
function of $\delta$, with an exponent of $1.0 \pm 0.01$ for all three systems
\cite{note}. This seems to be an interesting feature of all ferrimagnets.
The spin wave analysis however shows that this excitation gap is independent
of $\delta$ for the general ($s_1,s_2$) chain. The similar behaviors of these 
three alternating spin systems suggest that a ferrimagnet can be considered as
a ferromagnet with small quantum fluctuations. 

\section{Low-Temperature Properties}

In this section, we present results of our DMRG calculations of the
thermodynamic properties of the ($\frac{3}{2},1$), ($\frac{3}{2},\frac{1}{2}$) 
and ($1,\frac{1}{2}$) spin systems. The size of the system varies from $8$ to 
$20$ sites. We impose periodic boundary conditions to minimize finite size
effects with ${\vec S}_{1,N+1}$ = ${\vec S}_{1,1}$, so that the number of 
sites equals the number of bonds.
We set up the Hamiltonian matrices in the DMRG basis for all allowed 
$S_z$ sectors for a ring of $2N$ sites. We can diagonalize these matrices
completely to obtain all the eigenvalues in each of the $S_z$ sectors. 
As the number of DMRG basis states increases rapidly with 
increasing $m$, we retain a smaller number of dominant density matrix
eigenvectors in the DMRG procedure, $i.e.$, $50 \le m \le 65$, 
depending on the $S_z$ sector as well as the size of the system.
We have checked the dependence of properties (with $m$ in the
range $50 \le m \le 65$) for the system sizes we have studied 
($8 \le 2N \le 20$), and have confirmed that the properties do not
vary significantly for the temperatures at which they are computed;
this is true for all the three systems.
The above extension of the DMRG procedure is found to be accurate
by comparing with exact diagonalization results for small systems.
It may appear surprising that the DMRG technique which essentially
targets a single state, usually the lowest energy state in a chosen
sector, should provide accurate thermodynamic properties since these
properties are governed by energy level spacings and not the absolute
energy of the ground state. However, there are two reasons why the
DMRG procedure yields reasonable thermodynamic properties. Firstly,
the projection of the low-lying excited state eigenfunctions on the
DMRG space in which the ground state is obtained is substantial
and hence these excited states are well described in the chosen
DMRG space. The second reason is that the low-lying excitations
of the full system are often the lowest energy states in different
sectors in the DMRG procedure; hence their energies are quite accurate 
even on an absolute scale.

The canonical partition function $Z$ for the $2N$ site ring can be written as
\beq
Z ~=~ \sum_j ~e^{-\beta ( E_j - B(M)_j )} ~,
\eeq
where the sum is over all the DMRG energy levels of the
$2N$ site system in all the $S_z$ sectors. $E_j$ and $(M)_j$ denote the 
energy and the $\hat z$-component of the total spin
of the state $j$, $B$ is the strength of the magnetic field in units of
$J/g\mu_B$ ($g$ is the gyromagnetic ratio and $\mu_{B}$ is the
Bohr magneton) along the ${\hat z}$ direction, and $\beta=J/k_B T$ with 
$k_B$ and T being the Boltzmann constant and temperature respectively.
The field induced magnetization $< M >$ is defined as
\beq
< M > ~=~ \frac{\sum_j ~(M)_j ~e^{-\beta ( E_j - B(M)_j )}}{Z} ~.
\eeq
The magnetic susceptibility $\chi$ is related to the fluctuation in 
magnetization
\beq
\chi ~=~ \beta ~[~ < M^{2} > - < M >^{2} ~]~.
\eeq
Similarly, the specific heat $C$ is related to the fluctuation in the energy
and can be written as
\beq
C ~=~ \frac{\beta}{T} ~[~ < E^{2} > - < E >^{2} ~]~.
\eeq

The dimensionalities of the DMRG Hamiltonian matrices that we completely 
diagonalize vary from $3000$ to $4000$, depending upon the DMRG parameter 
$m$ and the $S_z$ value of the targetted sector, for rings of sizes greater
than $12$. These matrices are not very sparse, owing to the cyclic
boundary condition imposed on the system. The DMRG properties compare
very well with exact results for
small system sizes amenable to exact diagonalization studies. In the
discussion to follow, we present results on the $20$-site ring although all 
calculations have been carried out for system sizes from $8$ to $20$ sites.
This is because the qualitative behavior of the properties we have studied
are similar for all the ring sizes in this range for all three systems.

We present the dependence of magnetization on temperature for different
magnetic field strengths in figure 10 for all three systems. At low magnetic 
fields, the magnetization shows a sharp decrease at 
low temperatures and shows paramagnetic behavior at high temperatures. 
As the field strength is increased, the magnetization shows a slower decrease 
with temperature, and for high field strengths the magnetization shows 
a broad maximum. This behavior can be understood 
from the type of spin excitations present in these systems. The lowest
energy excitation at low magnetic fields is to a state with spin $s$
less than $s_G$. Therefore, the magnetization initially decreases at
low temperatures. As the field strength is increased, the gap to 
spin states with $s > s_G$ decreases as the Zeeman coupling to these
states is stronger than to the states with $s \le s_G$. The critical 
field strengths at which the magnetization increases with temperature
varies from system to system since this corresponds to the lowest spin gap
of the corresponding system. The behavior
of the system at even stronger fields turns out to be remarkable.
The magnetization in the ground state ($T=0$) shows an abrupt increase
signalling that the ground state at this field strength has $S_z >s_G$.
The temperature dependence of the magnetization shows a broad 
maximum indicating the presence of states with even higher spin values
lying above the ground state in the presence of this strong field. 
In all three cases, the ground state at very high field strengths
should be ferromagnetic. For the systems at such high fields,
the magnetization decreases slowly with increase of temperature as no 
other higher spin states lie above the ground state. While, we have not
studied such high field behaviors, we find that the field strength 
corresponding to switching the spin of the ground state $s_{G}$ to $s_{G}+1$ 
is higher for ($\frac{3}{2},\frac{1}{2}$) system compared to ($\frac{3}{2},1$)
and ($1, \frac{1}{2}$) systems. The switching field appears to
depend on the value of $|s_1 -s_2 |$. We see in figure 10 that in the
($\frac{3}{2},1$) and ($1,\frac{1}{2}$) cases, the ground state
has switched to the higher spin state at highest magnetic field strength
we have studied but in the ($\frac{3}{2},\frac{1}{2}$) case, the ground
state has not switched even at the field strength indicating that the
excitation gap for this system is the larger than the other two. For the 
($\frac{3}{2},\frac{1}{2}$) case, the 
same situation should occur at very high magnetic fields. Thus, we predict
that the highest $s_z$ is attained in the ground state at high
magnetic field and this field strength increases with increase in
site spin difference $|s_1 - s_2|$.

The dependence of magnetization on the magnetic field is shown at different 
temperatures in figure 11 for all three systems studied here. At low 
temperature the magnetization shows a plateau. The width of the plateau 
depends on the system and it decreases as the temperature is raised. 
Eventually the plateau disappears at higher temperatures. The existence of
the plateau shows that the higher spin states are not accessible 
at the chosen temperature. The plateau
is widest in the ($\frac{3}{2},\frac{1}{2}$) case once again reflecting the 
larger gap to the $s_{G}+1$ state in this system compared with the 
($\frac{3}{2}-1$) and ($1, \frac{1}{2}$) systems. At higher fields, the larger 
Zeeman splittings of higher spin states become accessible leading to an
increase in the magnetization. The magetization curves 
at all temperatures intersect at some field strengths depending on the
system. For the ($\frac{3}{2},1$) case, the intersection occurs at 
$B=1.0J/g\mu_B$ and at higher field strength all the curve collapse.
For ($\frac{3}{2},\frac{1}{2}$) case, the intersection occurs 
at $B=1.5J/g\mu_B$, while for the ($1,\frac{1}{2}$) system
these curves intersect twice, once at $B=1.0J/g\mu_B$ and again at 
$B=2.5 J/g\mu_B$ in the chosen field range. These fields are close to 
the field strengths at which the ground state switches from one 
$S_z$ value to a higher value for the corresponding system. Thus, they
are strongly dependent upon the actual values of $s_1$ and $s_2$.

The dependence of $\chi T/2N$ on temperature for different
field strengths are shown in figure 12 for all three systems. For 
zero field, the zero temperature value of $\chi T$
is infinite in the thermodynamic limit; for finite rings it is 
finite and equal to the average of the square of the magnetization in
the ground state. For the ferrimagnetic ground state $\chi T/2N$, as $T 
\rightarrow 0$, is given by $s_G (s_G +1)/6N$. As the 
temperature increases, this product decreases and shows a minimum
before increasing again. For the three systems studied here, the minimum
occurs at different temperatures depending on the system. 
For the ($\frac{3}{2},1$) alternating spin system, it is at $k_B T=(0.8 \pm 
0.1)J$, while for the ($\frac{3}{2},\frac{1}{2}$) and ($1,\frac{1}{2}$) cases, 
it occurs at $k_B T=(1.0 \pm 0.1)J$ and $k_B T=(0.5 \pm 0.1)J$ respectively. 
The minimum occurs due to the states with $S_z < s_G$ getting populated at 
low temperatures. In the infinite chain limit, these states turn out
to be the gapless excitations of the system.  The subsequent
increase in the product $\chi T$ is due to the higher energy-higher spin 
states being accessed with further increase in temperature.
This increase is slow in ($\frac{3}{2}, \frac{1}{2}$) case, as in this 
system very high spin states are not accessible within the chosen temperature 
range. Experimentally, it has been found in the bimetallic chain compounds
that the temperature at which the minimum occurs in the $\chi T$ 
product depends upon the magnitude of the spins $s_1$ and $s_2$ \cite{kahn3}.
The $Ni^{II}-Cu^{II}$ bimetallic chain shows a minimum
in $\chi T/2N$ at a temperature corresponding to $55$ $cm^{-1}$ ($80 K$);
an independent estimate of the exchange constant in this system is
$100$ $cm^{-1}$ \cite{kahnpc}. This is in very good agreement with the
minimum theoretically found at temperature $(0.5 \pm 0.1)J$ for the 
($1,\frac{1}{2}$) case. The minimum in $\chi T/2N$ vanishes at 
$B=0.1 J/g\mu_B$ which corresponds to about $10 T$ for all three systems. It
would be interesting to study the magnetic susceptibility of these
systems experimentally under such high fields. The low-temperature 
zero-field behavior of our systems
can be compared with the one-dimensional ferromagnet. In the
latter, the spin wave analysis shows that the $\chi T$ 
product increases as $1/T$ at low temperatures \cite{taka}. 

In finite but weak fields, the behavior of $\chi T$ is different. 
The magnetic field opens up a gap and $\chi T$ falls exponentially 
to zero for temperatures less than the gap in the applied field for all
three systems. Even in this case a minimum is found at the same temperature 
as in the zero-field case for the corresponding system, for the same reason 
as discussed in the zero field case.

In stronger magnetic fields, the behavior of $\chi T$ 
from zero temperature upto $k_B T=J_{min}$ ($J_{min}$ is the temperature
at which the minimum in $\chi T$ is observed) is qualitatively different. The 
minimum in this case vanishes for all three systems. In these field strengths, 
the states with higher $S_z$ values are accessed even below $k_B T=J_{min}$.
The dependence of $\chi T$ above $k_B T=J_{min}$ at all field strengths is 
the same in all three systems.
In even stronger magnetic fields, the initial sharp increase is
suppressed. At very low temperature, the product $\chi T$ is nearly
zero and increases almost linearly with $T$ over the temperature range
we have studied. This can be attributed to a switch in the ground state
at this field strength. The very high temperature behavior of $\chi T$
should be independent of field strength and should saturate to the
Curie law value corresponding to the mean of magnetic moments due to 
spin-$s_1$ and spin-$s_2$.

The temperature dependence of specific heat also shows a marked
dependence on the magnetic field at strong fields. This dependence
is shown in figure 13 for various field strengths for all the three systems. 
In zero and weak magnetic fields, the specific heat shows a broad maximum at 
different temperatures which are specific to the system. Interestingly, the 
temperature at which the specific heat shows a maximum closely corresponds to 
the temperature where the low-field $\chi T$ has a minimum for the 
corresponding system. For a strong magnetic field ($B=J$), there is a dramatic
increase in the peak height at about the same temperature corresponding
to the specific system, although
the qualitative dependence is still the same as at low magnetic fields
in all three cases. This phenomena indicates that the higher energy 
high-spin states are brought to within $k_B T$ of the ground state 
at this magnetic field strength for all three cases.

Studies of thermodynamic properties of the dimerized alternating spin chains 
in these three cases show qualitatively similar trends to that of the 
corresponding uniform systems and follows from the fact that the low-energy 
spectrum does not change qualitatively upon dimerization.

\section{Summary}

We have studied alternating spin-${\frac{3}{2}}$/spin-$1$ and
spin-$\frac{3}{2}$/spin-$\frac{1}{2}$ systems in detail and compared them
with our earlier studies on spin-1/spin-$\frac{1}{2}$ model. The ground state
and low-lying excited states have been analyzed by employing the DMRG method
and have been compared with the corresponding spin-wave results. The spin
of the ground state is given by $s_G =N(s_1 -s_2)$ for a $2N$ site system in 
all the three cases. There are two types of spin excitations in all the 
systems. The lowest excitation one is gapless in the infinite chain limit to 
a state with spin $s_G -1$. Another excitation to the state with spin $s_G +1$ 
is gapped and the gap is larger when the
difference in site spins $|s_1 - s_2|$ is an integer. Interestingly,
the low-energy spectrum remains qualitatively unchanged upon dimerization,
and the dimerization is itself conditional in the infinite chain limit 
for all the three systems.

We have also employed the DMRG technique to obtain low-temperature
thermodynamic properties. For all these cases, the magnetic susceptibility
shows very interesting magnetic field dependence. The $\chi T$ vs. $T$
plot shows a minimum at low-magnetic fields, and the minimum vanishes
at high-magnetic fields. The temperature at which this minimum 
occurs varies from system to system. The specific heat shows a maximum
as a function of temperature at all fields. Moreover, the height
of the maximum shows a dramatic increase at high-magnetic field.
Interestingly, in each system, the temperature corresponding to the 
maximum in $C$ and minimum in $\chi T$ is the same at low-magnetic fields. 

{\bf Acknowledgements}

We thank Professor Olivier Kahn who introduced one of us (S.R.) to the 
experimental alternating spin systems and therby motivated us to undertake
this work. The present work has been supported by the 
Indo-French Centre for the Promotion of Advanced Research through project 
No. 1308-4, "Chemistry and Physics of Molecule based Materials".

\vskip 1 true cm

\newpage

{\centerline{\bf Figure Captions}} 
\vskip .5 true cm

\noindent {\bf Figure 1} \\
Extrapolation of the ground state energy per site ($\epsilon_0$), in units of
$J$, as a function of inverse system size.

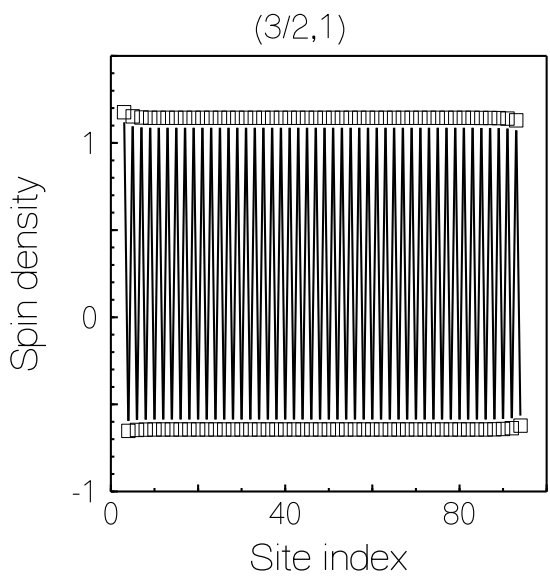
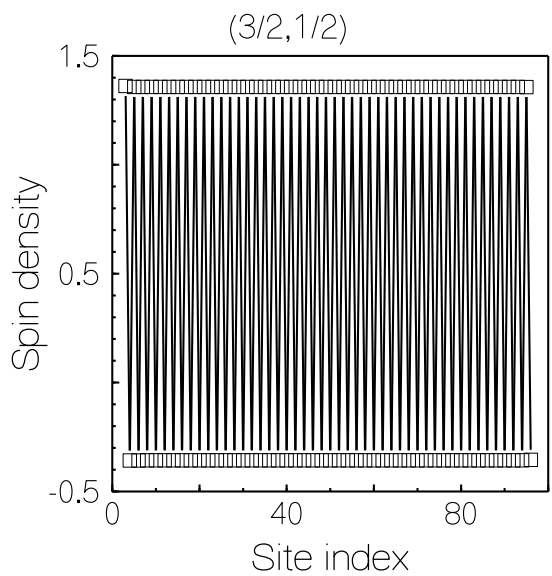
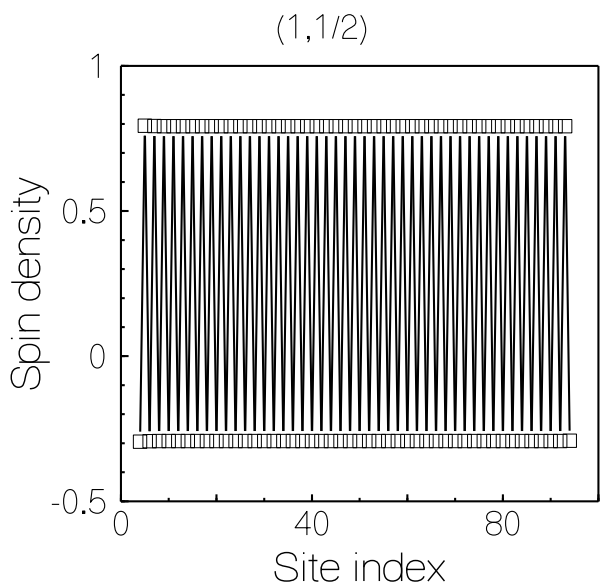
\noindent {\bf Figure 2} \\
Expectation values of the $z$-components of the two spins vs. the unit cell 
index $n$. The upper and lower points are for the spin-$s_1$ and the 
spin-$s_2$ sites respectively.

\noindent {\bf Figure 3} \\
Subtracted two-spin correlation functions as a function of distance between 
the two spins. (a) spin-$s_1$ spin-$s_1$ correlations, (b) spin-$s_2$ 
spin-$s_2$ correlations, and (c) spin-$s_1$ spin-$s_2$ correlations. In each 
figure, squares correspond to ($\frac{3}{2},1)$), circles to 
($\frac{3}{2},\frac{1}{2}$) and triangles to ($1,\frac{1}{2}$) systems.

\noindent {\bf Figure 4} \\
Energy difference (units of $J$) between the ground state and the lowest 
energy state with spin $s=s_G -1$ as a function of inverse system size. 
$s_G$ is the total spin of the ground state.

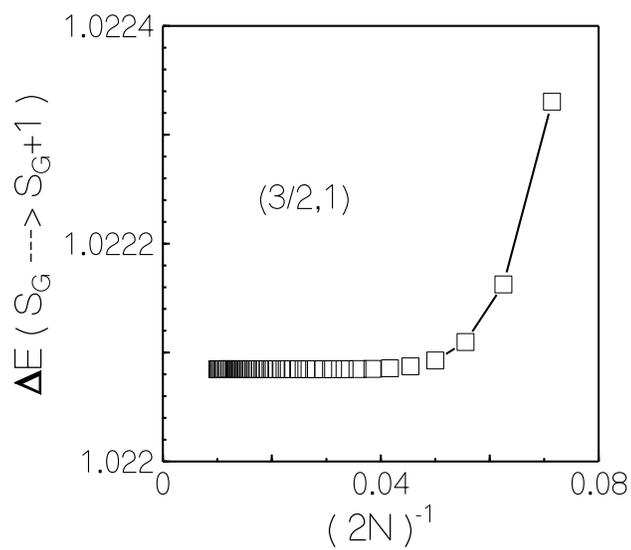
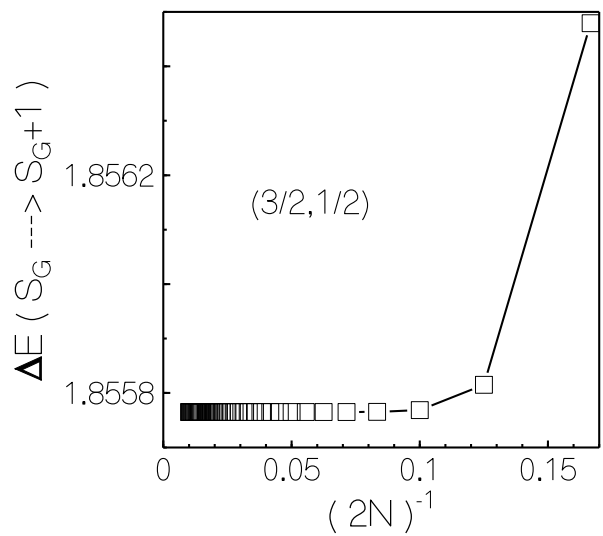
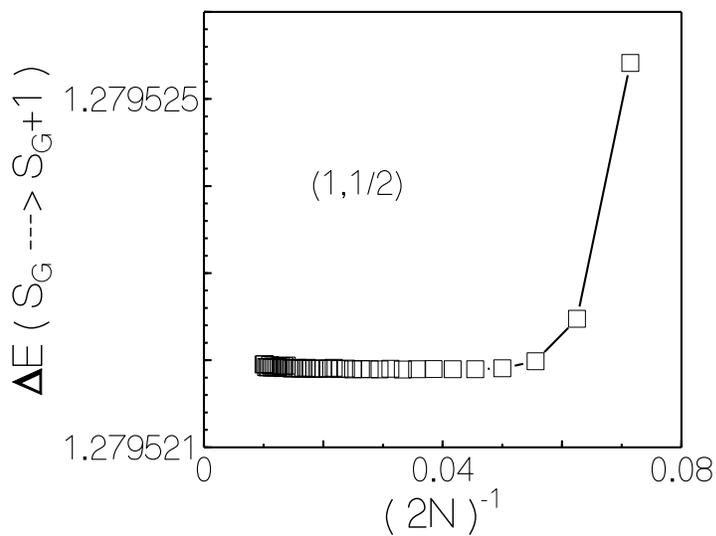
\noindent {\bf Figure 5} \\
Excitation gap (units of $J$) from the ground state (spin $s=s_G$) to the
state with spin $s=s_G +1$ as a function of the inverse system size. 

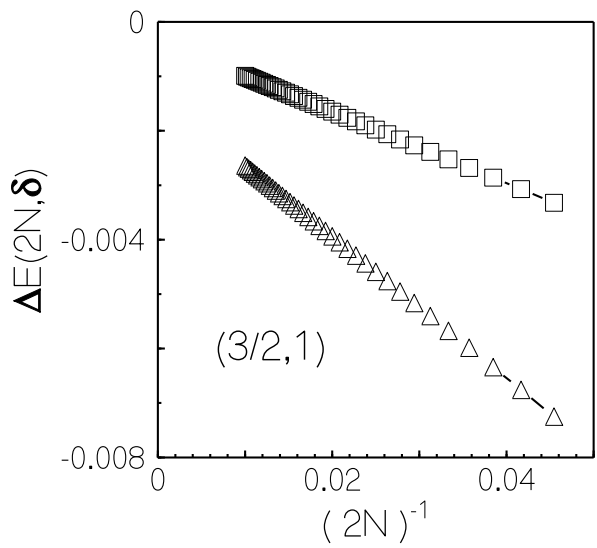
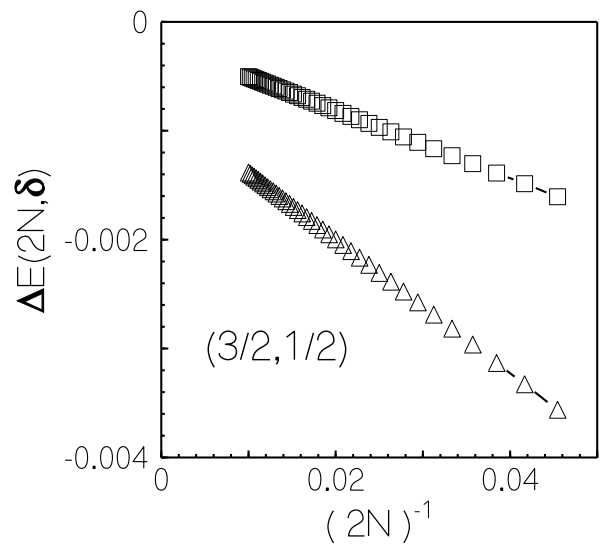
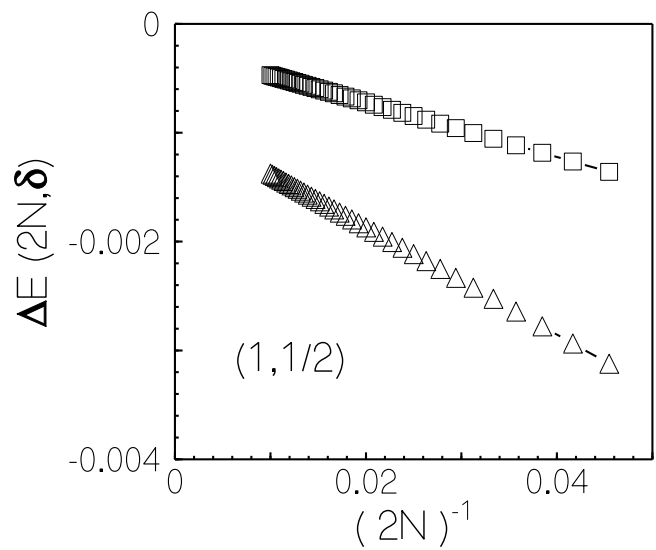
\noindent {\bf Figure 6} \\
Gain in magnetic energy $\Delta E(2N, \delta)$ (units of $J$) associated with 
dimerization vs. the inverse system size for two different values of 
dimerization $\delta$. $\delta = 0.025$ (squares) and $\delta = 0.05$ 
(triangles).

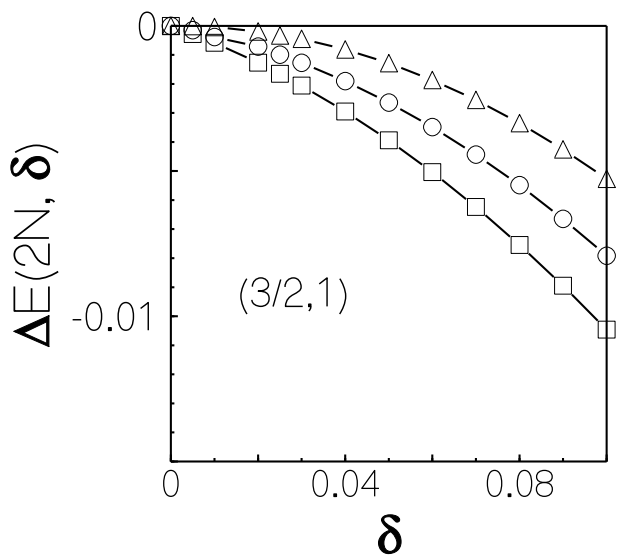
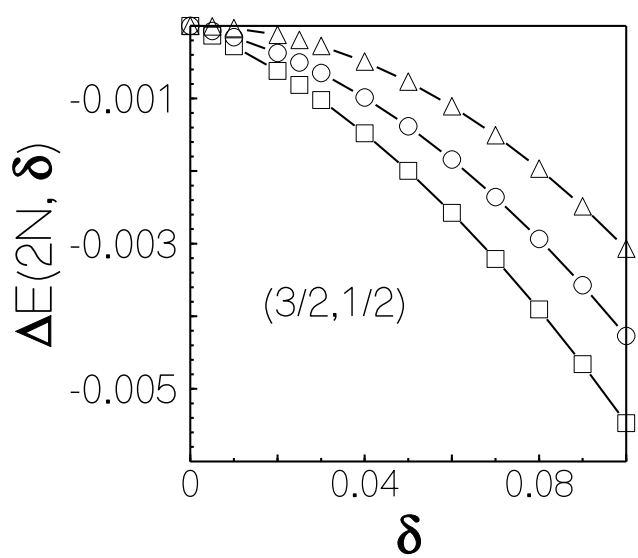
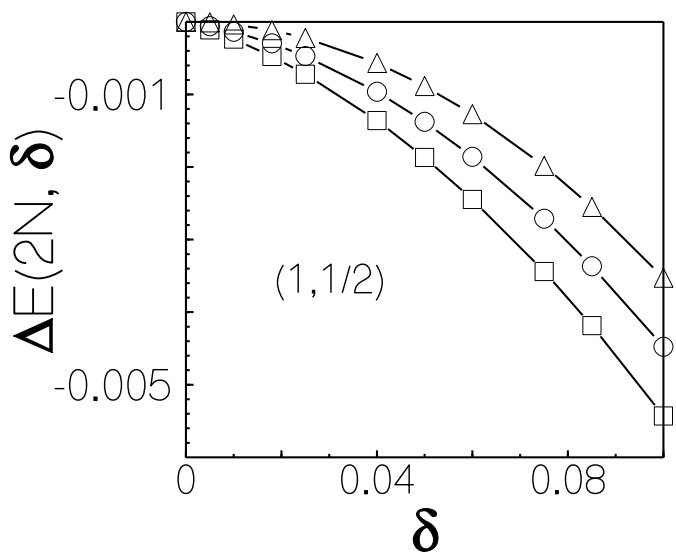
\noindent {\bf Figure 7} \\
Magnetic energy gain (units of $J$) as a function 
of dimerization parameter $\delta$ for different system sizes. In the figure 
$2N= 50$ (squares), $2N= 100$ (circles) and extrapolated values with
$N \rightarrow \infty$ (triangles) are shown.

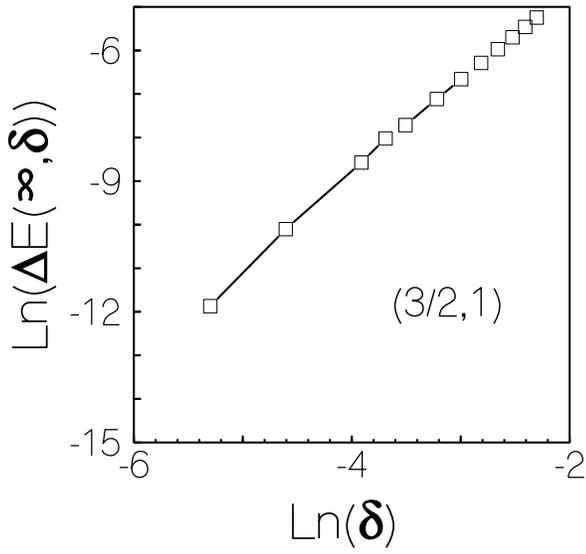
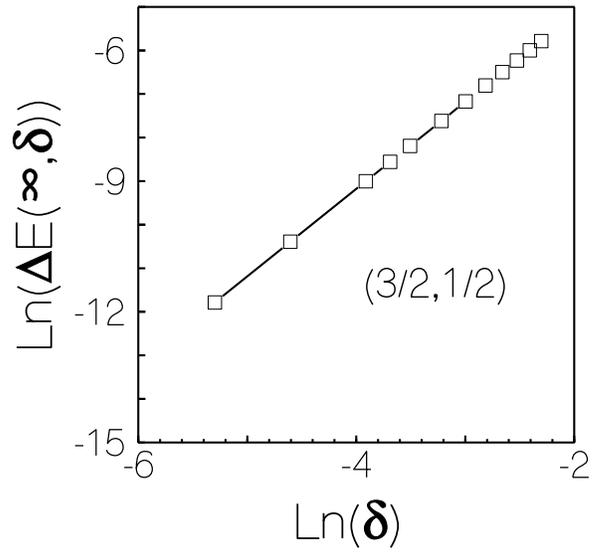
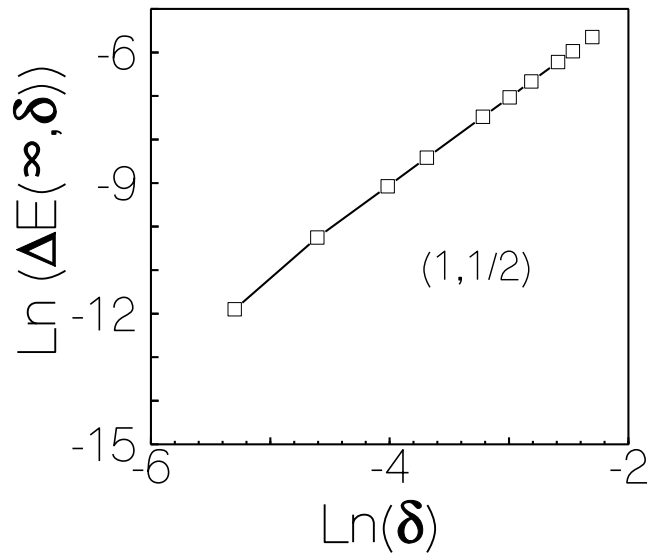
\noindent {\bf Figure 8} \\
Log-log plot of extrapolated magnetic energy gain (units of $J$) for infinite 
system size and dimerization parameter $\delta$. Interestingly, the slope 
is found to be $2.00 \pm 0.01$ for all the three systems.

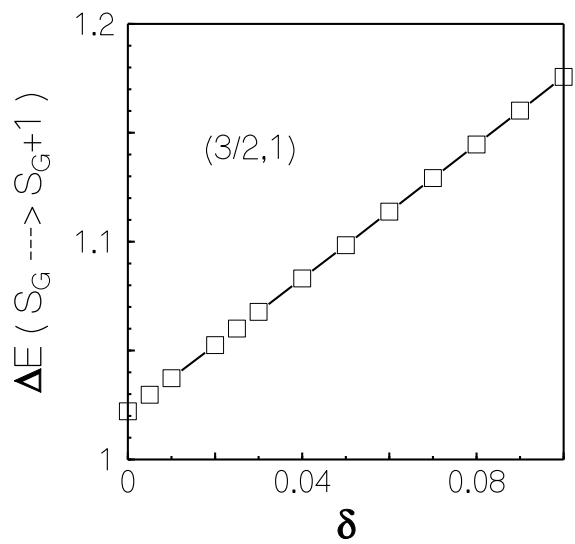
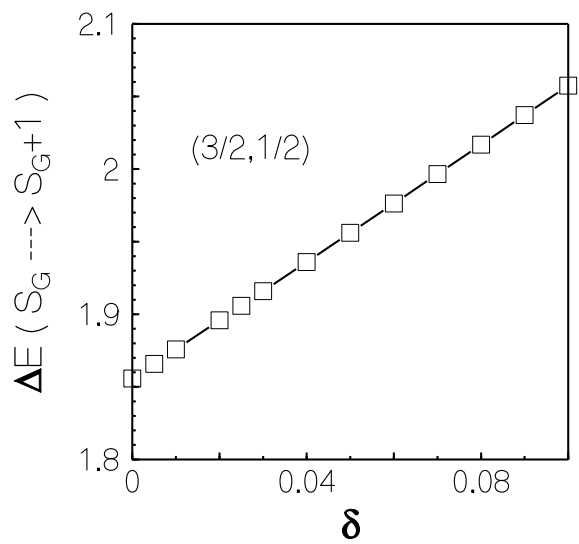
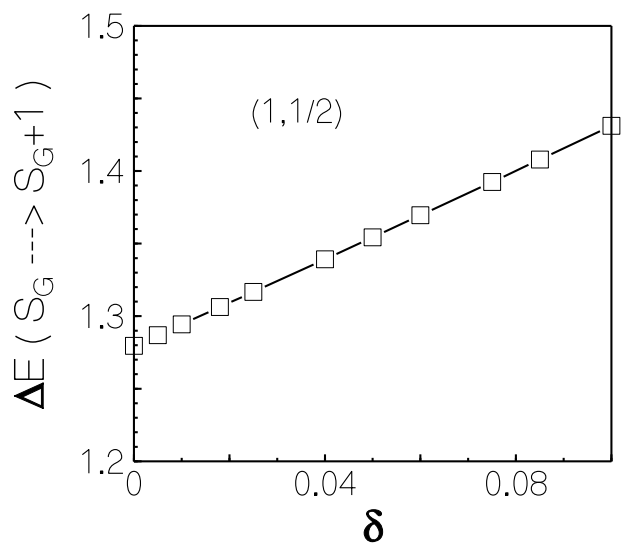
\noindent {\bf Figure 9} \\
Excitation gap (units of $J$) to the state with spin $s=s_G +1$ from the 
ground state ($s=s_G$) as a function of $\delta$ for the dimerized 
alternating chain. The exponent is $1.0 \pm 0.01$ for all three systems.

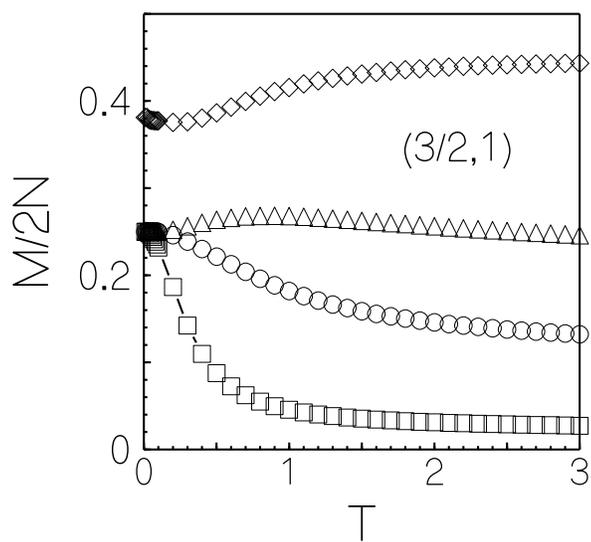
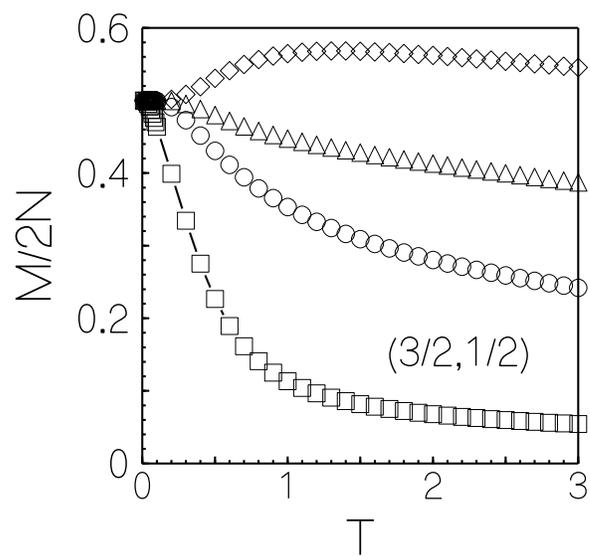
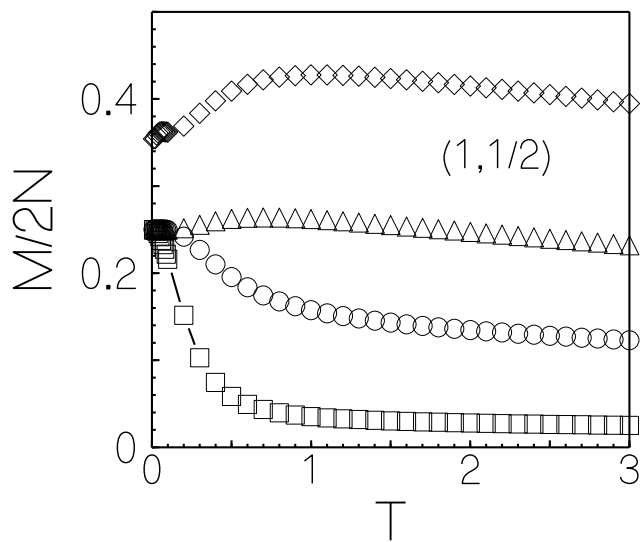
\noindent {\bf Figure 10} \\
Plot of magnetization per site as a function of temperature $T$ for four
different values of the magnetic fields $B$. Squares are for $B=0.1 
J/g\mu_{B}$, circles for $B=0.5 J/g\mu_{B}$, triangles for $B=J/g\mu_{B}$ 
and diamonds for $B=2 J/g\mu_{B}$.

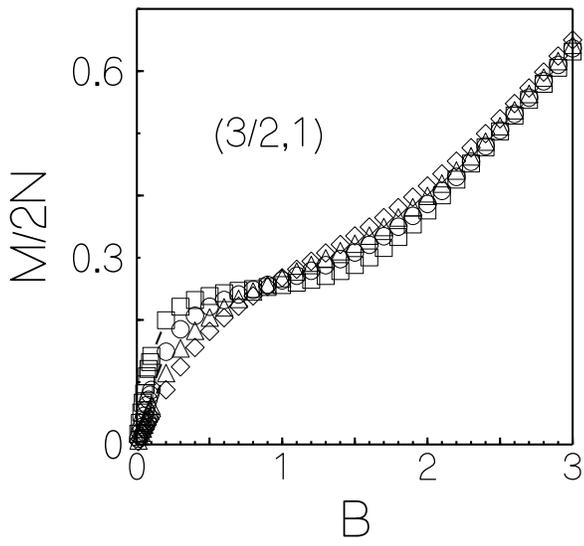
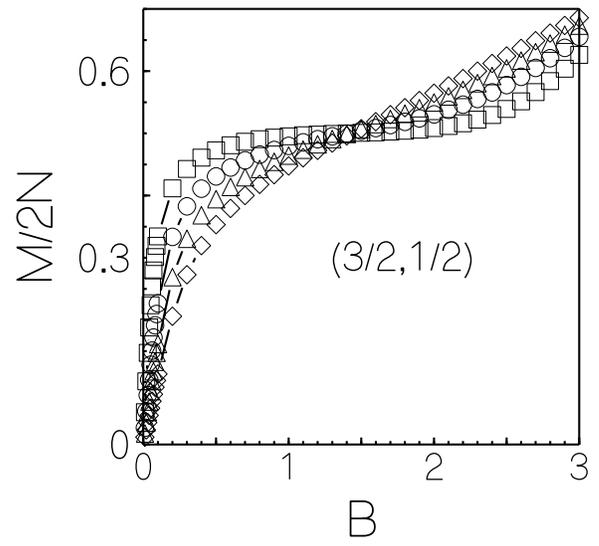
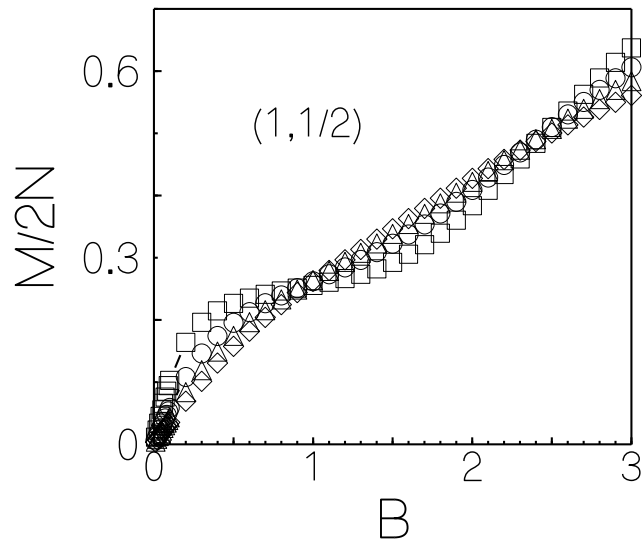
\noindent {\bf Figure 11} \\
Magnetization per site vs. the magnetic field strength $B$, in units of 
$J/g\mu_{B}$, for four different temperatures $T$. $T=0.3 J/k_{B}$ results are 
given by squares, $T=0.5 J/k_{B}$ by circles, $T=0.7 J/k_{B}$ by triangles and 
$T=J/k_{B}$ by diamonds.

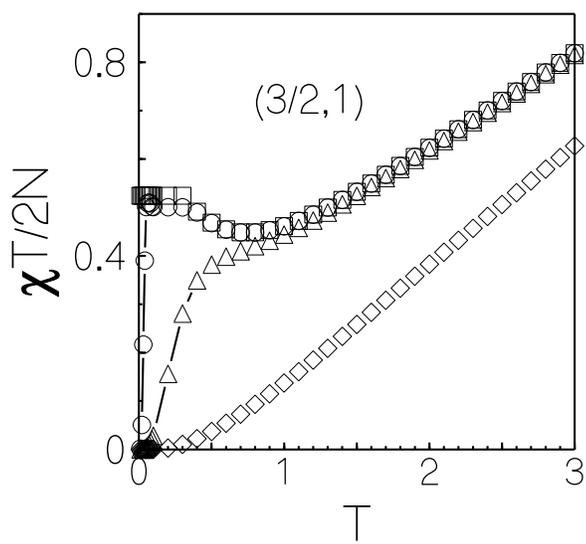
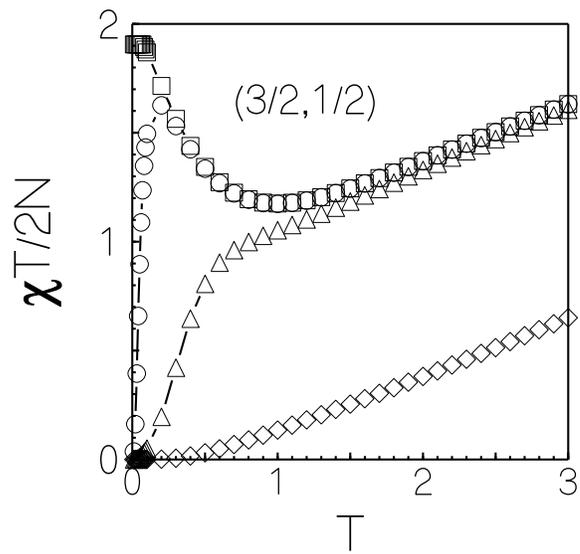
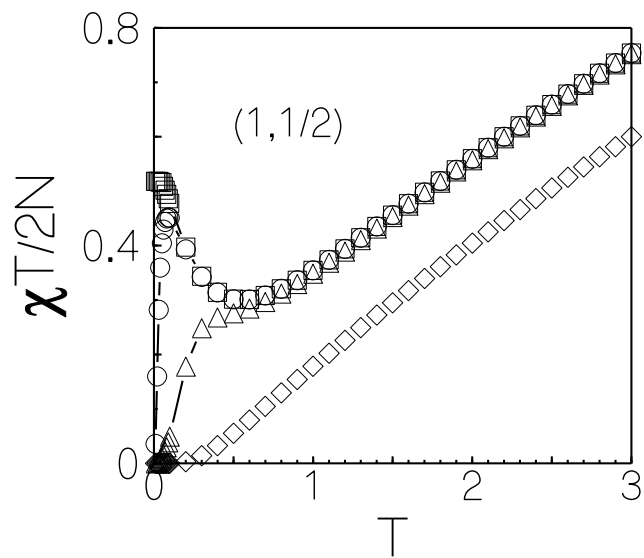
\noindent {\bf Figure 12} \\
$\chi T$ ( defined in the text) per site as a function of temperature $T$
for various magnetic fields $B$. Zero field results are shown by squares, 
$B=0.01 J/g\mu_{B}$ by circles, $B=0.1 J/g\mu_{B}$ by triangles and 
$B=J/g\mu_{B}$ by diamonds.

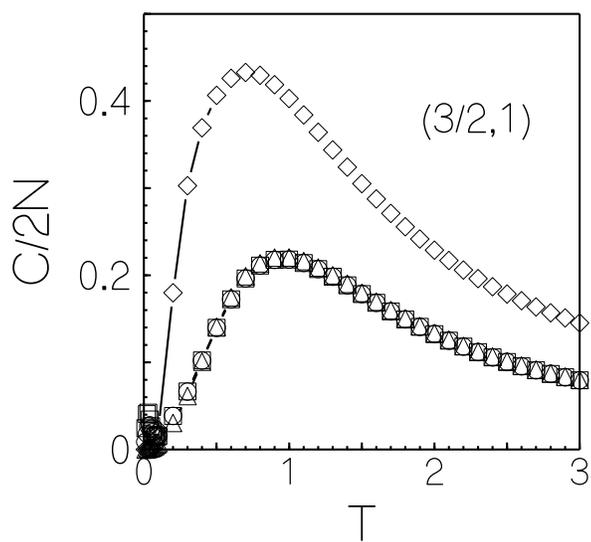
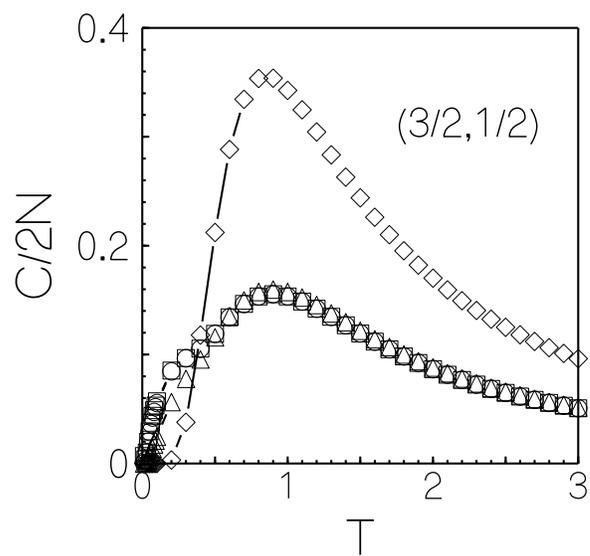
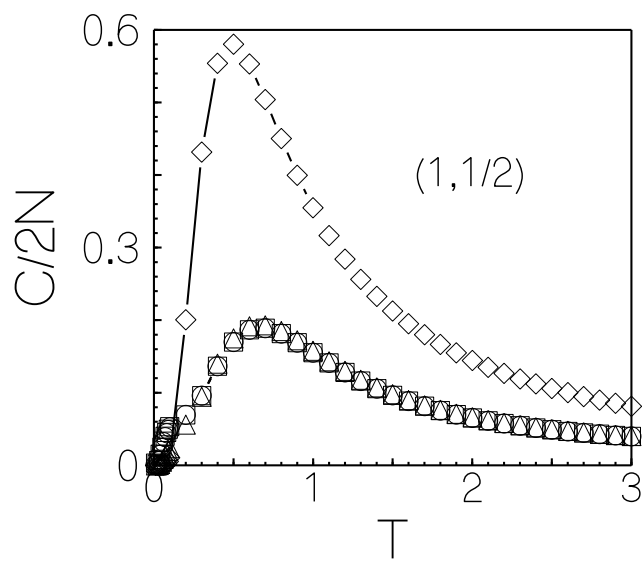
\noindent {\bf Figure 13} \\
Specific heat per site as a function of temperature $T$ for four different
values of magnetic fields $B$. Zero field data are shown by squares, $B=0.01 
J/g\mu_{B}$ by circles, $B=0.1 J/g\mu_{B}$ by triangles and $B=J/g\mu_{B}$ by 
diamonds.


\begin{thebibliography}{99}
\bibitem{note} Earlier extrapolation in the ($1, \frac{1}{2}$) case was done 
over the range $\delta$ $\epsilon$ $0$ to $0.6$. Present extrapolations are 
for the range $\delta$ $\epsilon$ $0$ to $0.1$ and hence a small difference 
from the earlier quoted value of $1.07$ in reference \cite{pati}.
\end{thebibliography}
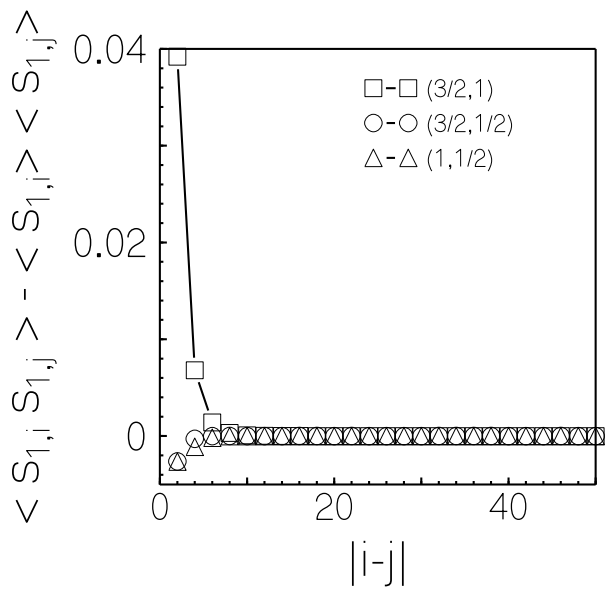
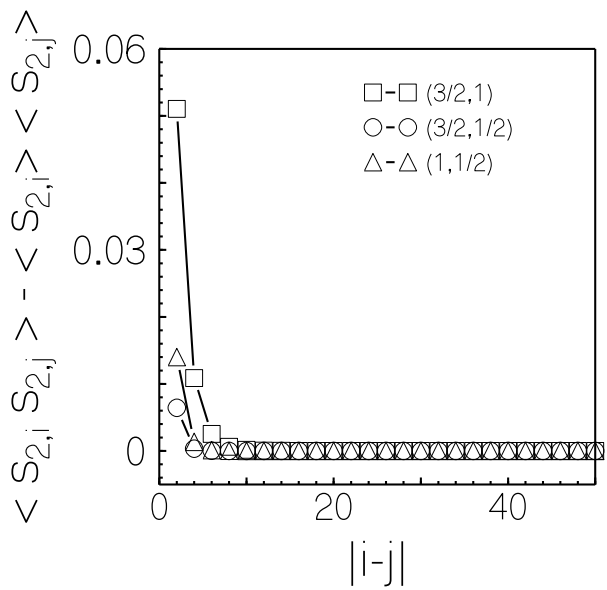
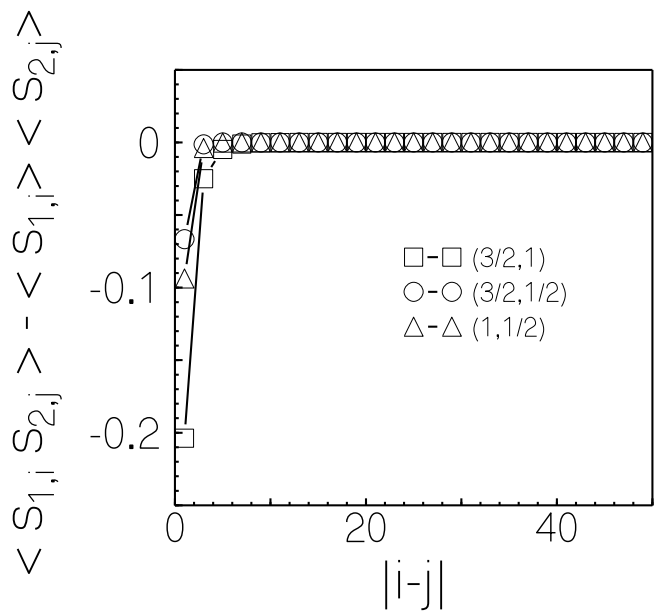
\end{document}